\documentclass[graybox]{svmult}

\usepackage{mathptmx}       
\usepackage{helvet}         
\usepackage{courier}        
\usepackage{type1cm}        
\usepackage{makeidx}         
\usepackage{graphicx}        
\usepackage{multicol}        
\usepackage[bottom]{footmisc}

\usepackage{amsmath,amssymb,amsfonts}
\usepackage{bbold}

\makeindex             

\begin{document}

\title*{Transport in topological insulator nanowires}
\author{Jens H. Bardarson and Roni Ilan}
\institute{Jens H. Bardarson \at Department of Physics, KTH Royal Institute of Technology, Stockholm, SE-106 91 Sweden, \email{bardarson@kth.se}
\and Roni Ilan \at Raymond and Beverly Sackler School of Physics and Astronomy, Tel-Aviv University, Tel-Aviv 69978, Israel, \email{ronilan@tauex.tau.ac.il}}
%
%
\maketitle

\abstract{
In this chapter we review our work on the theory of quantum transport in topological insulator nanowires. 
We discuss both normal state properties and superconducting proximity effects, including the effects of magnetic fields and disorder.
Throughout we assume that the bulk is insulating and inert, and work with a surface-only theory.
The essential transport properties are understood in terms of three special modes: in the normal state, half a flux quantum along the length of the wire induces a perfectly transmitted mode protected by an effective time reversal symmetry; a transverse magnetic field induces chiral modes at the sides of the wire, with different chiralities residing on different sides protecting them from backscattering; and, finally, Majorana zero modes are obtained at the ends of a wire in a proximity to a superconductor, when combined with a flux along the wire. 
Some parts of our discussion have a small overlap with the discussion in the review~\cite{Bardarson:2013cn}.
We do not aim to give a complete review of the published literature, instead the focus is mainly on our own and directly related work. 
}
\section{Overview and general considerations}
\label{sec:1}

Topological insulators (TI's)~\cite{Hasan:2010ku,Qi:2011hb,Hasan:2011hs} are characterized by their bulk-boundary correspondence: the bulk has a gap that is inverted in comparison with the atomic insulator (vacuum), resulting in a robust metallic state at the surface.
The quantum Hall effect, with its Landau levels and chiral edge states, is a good example. 
In this case the Hamiltonian has no symmetries (apart from charge conservation) and quantum Hall states are realized in even spatial dimensions.
The presence of symmetries allows for symmetry-protected topological phases~\cite{Chiu:2016et}, as long as the symmetry is not broken.
The quantum spin Hall effect in 2D is the time reversal invariant version of the quantum Hall effect and the metallic surface consists of two counter-propagating helical edge states that are Kramers pairs and therefore not coupled by time reversal preserving disorder~\cite{Kane:2005gb,Kane:2005hl,Bernevig:2006ij}.
Particle-hole symmetry allows for topological superconductivity in which case the surface states are particle-hole symmetric Majorana zero modes~\cite{Kitaev:2001gb,Fu:2008gu}.

In this review we focus on 3D topological insulators protected by time reversal symmetry~\cite{Hasan:2011hs}.
In this case the surface is 2D and the low energy degrees of freedom comprise an odd number (which we take to be one) of Dirac fermions.
A defining feature of topological insulators is reflected in the fact that a single Dirac fermion can not be localized, no matter how strong the disorder~\cite{Bardarson:2007iu,Nomura:2007jb}.
Instead, disorder always drives the surface in the thermodynamic limit into a metallic phase referred  to as the symplectic metal~\cite{Ostrovsky:2007hc,Ryu:2007bu}.
Interference in the symplectic metal gives rise to weak anti-localization~\cite{Hikami:1980jn}, the phenomena that the lowest order quantum correction to the classical Drude conductivity is positive, leading to an enhanced conductance. 
While one can understand this as being destructive interference of time reversed loops due to the Berry phase picked up by the Dirac fermion as it loops around, enforced by spin-momentum locking, it is not a signature of topology---any 2D strongly spin-orbit coupled metal is, ignoring interactions, symplectic.

This Berry phase and the time-reversal symmetry strongly affect transport properties and are the key effects in the physics we discuss in this chapter.
In the presence of a time-reversal symmetry $\mathcal{T}$ that satisfies $\mathcal{T}^2= -1$, the scattering matrix $S$ that relates incoming modes to outgoing modes in a two terminal scattering setup, is antisymmetric: $S^T = -S$~\cite{Bardarson:2008jk}.
As a consequence, backscattering is forbidden (the diagonal elements of the scattering matrix are zero) and in the presence of an odd number of modes, a perfectly transmitted mode~\cite{Ando:1998fn} with transmission unity, is obtained. 
In the presence of a perfectly transmitted mode, the conductance, via the Landauer formula, $G \geq e^2/h$, and localization can not take place.
In the field theory of diffusion this is encoded in the presence of a topological term~\cite{Ostrovsky:2007hc,Ryu:2007bu}; all topological insulators and superconductors can in fact be classified according to the presence or absence of a topological term in the corresponding non-linear sigma model describing diffusion~\cite{Schnyder:2008ez,Ryu:2010ko}.

In the limit of large number of modes, and conductance $G \gg e^2/h$, the distinction between and odd and even number of modes is not important. 
This observation has been used to argue for the absence of localization in weak topological insulators~\cite{Ringel:2012hz,Mong:2012du}.
Here, it means that in the thermodynamic limit, transport can not distinguish a 3D topological insulator surface from any regular spin-orbit coupled metal, and no direct signatures of topology are to be obtained.
This is the main motivation for exploring the transport properties of topological insulator nanowires.
By reducing the size of the surfaces, the distinction between an even and odd number of modes becomes important and a direct signature of topology can be obtained in the presence of a perfectly transmitted mode at the Dirac point, which results in a quantized conductance of $e^2/h$~\cite{Bardarson:2010jl}.
The perfectly transmitted mode requires magnetic flux along the length of the wire~\cite{Ostrovsky:2010eq,Rosenberg:2010dj,Bardarson:2010jl,Zhang:2010bd}
A transverse magnetic field induces quantum Hall phases at the top and bottom of a wire, with chiral modes at the sides~\cite{Lee:2009do}.
These modes and their essential transport properties are discussed in Sec.~\ref{sec:2}.

The odd number of modes is especially important when it comes to superconducting proximity effect: an s-wave superconductor coupled to a Dirac fermion with an odd number of modes can result in topological superconductivity with Majorana zero modes~\cite{Fu:2008gu,Cook:2011ij}.
Such a topological superconducting wire, when coupled with the perfectly transmitted or chiral mode of the normal state, has distinct transport signatures.
Due to B{\'e}ri degeneracy~\cite{Beri:2009kj}, the NS conductance of a normal metal -- superconductor interface, in the single mode limit, is either $0$ or $e^2/h$~\cite{Wimmer:2011fq}.
A magnetic flux along the wire allows to turn the topological superconductivity on and off. 
These and other related superconducting transport phenomena are discussed in Sec.~\ref{sec:3}.

\section{Topological insulator nanowires: normal state properties}
\label{sec:2}

Topological insulator nanowires come in many shapes and sizes~\cite{Peng:2010jm,Xiu:2011hq,Dufouleur2013a,Hong2014,Cho:2015gk,Jauregui:2016cx}.
Their cross sections are commonly rectangular and the wires look like ribbons.
Their bulk is frequently inescapably doped during synthesis, and is far from being an ideal insulator.
Nevertheless, the surface is a significant contributor to transport, and often the most characteristic features of experimental data are surface features~\cite{Hamdou:2013hb}.
That, in addition to rapid improvements in the material science and synthesis of wires with more and more insulating bulks, motives us to make the simplifying theoretical assumptions of inert insulating bulk.
The metallic surface state is modeled by a single Dirac fermion, which is sharply localized at the surface.
By taking into account the effects of disorder, doping and magnetic field on transport of such a surface state, this theory has proven to be sufficiently detailed to describe the essential features of most experiments.

What are the defining properties of a wire? 
The most important feature is the aspect ratio of circumference $P$ and the length $L$, which we usually take to be of the order of unity or smaller.
In the limit $P/L \gg 1$, transport is independent of boundary conditions~\cite{Tworzydio:2006hw}, and therefore magnetic flux, and the conductance flows into the symplectic metal~\cite{Bardarson:2007iu,Nomura:2007jb}.
A typical circumference is of order 100 $nm$, which is a couple of orders of magnitude larger than a typical carbon nanotube.
This is has two important consequences: first, the magnetic field strengths needed to thread a flux quantum through the wire is easily realized experimentally, and second, the energy scale of confinement in the transverse direction $\Delta_P = \hbar v_F 2\pi/P$ is small enough that the number of modes can be tuned by gating, while at the same time temperature can be lowered such that individual modes can be resolved.
Similarly, disorder broadening of transverse modes $\Gamma < \Delta_P$~\cite{Dufouleur:2017de}.
The transport regime is therefore quasi-one dimensional, with typically multiple modes taking part in transport but separate transverse modes being resolvable.

In this quasi-1D limit, signatures of topology are visible in transport properties.
These features, relying on topology, are insensitive to the detailed geometry of the wires. 
Most of the time, no significant qualitative changes are observed in the results of transport calculations if one assumes the wires to be perfectly cylindrical instead of having the more realistic rectangular shapes. 
This assumption simplifies notation and some calculations and we therefore often make it.
A magnetic field transverse to the length of the wire, however, breaks rotational symmetry; a cylindrical shape no longer leads to simplifications and we revert to rectangular shapes.

\subsection{Band structure of a clean wire}
\label{sec:bands_normal}

In a compact geometry, such as that of a nanowire, the Berry phase due to spin-momentum locking leaves its hallmarks on the electronic band structure. 
The spin of a Dirac fermion is locked to the momentum direction and therefore rotates as the momentum goes in a loop.
This is what happens when the Dirac fermion encircles the circumference of a nanowire.
As a result, with a $2\pi$ rotation of a spin giving a minus sign, the wave function is antiperiodic~\cite{Ran:2008dv,Rosenberg:2010dj}.
Alternatively one can keep the wave function periodic and include a spin connection term in the Hamiltonian~\cite{Zhang:2009fu,Ostrovsky:2010eq}; these description are equivalent up to a gauge transformation. 
Antiperiodicity requires nonzero transverse momentum, and therefore energy, necessitating a gap in the energy spectrum.

We demonstrate this for a simple model of a topological insulator surface state, with the effective Hamiltonian of a single Dirac fermion living on the surface of a cylindrical wire of circumference $P$. 
The surface Hamiltonian is given by (we set $\hbar = 1$)
\begin{equation}
H = -iv_F\left[
\sigma_x \partial_x +  2\pi/P\sigma_y \partial_s  
\right]    \label{H_wire}
\end{equation}
which is equivalent to the Hamiltonian of a Dirac fermion in a flat surface, with he crucial difference of antiperiodic boundary condition on the wave function in the compact coordinate $0\leq s\leq P$: $\psi(s+P) = -\psi(s)$. 
$v_F$ is the Fermi velocity, $\sigma_x$ and $\sigma_y$ the Pauli spin matrices. 
The wave functions on the cylinder take the form $$\psi_{k,n}=e^{ikx+i\ell_ns}\chi_{k,n},$$ with $\ell_n=n-1/2$ and $n\in {Z}$.  
The spinor $\chi_{k,n}$ satisfies $\hat{p}\cdot\sigma \chi_{k,n} = \pm \chi_{k,n}$, with $\hat{p}$ the unit vector in the direction of the momentum.
The band structure 
\begin{equation}\label{bands}
E_{k,n}=\pm v_F\sqrt{k^2+(2\pi/P)^2\ell_n^2}.
\end{equation}
is gapped with a finite gap of magnitude $\Delta_P=2\pi v_F/P$ at $k=0$.
All energy bands are doubly degenerate.

We emphasize that the Hamiltonian given in Eq.~\eqref{H_wire} is in fact suitable to account for the physics of the surface states of wires with any (constant) cross-section, provided they are strictly two dimensional and uniform. 
Nevertheless, as already mentioned, realistic systems do not necessarily meet these requirements. 
For example, they may have slightly different effective Dirac Hamiltonian depending on the surface termination.
The band structure of rectangular wires has been studied both analytically and numerically taking into consideration  corrections due to such details~\cite{Brey:2014db}. 
The result is qualitatively the same as obtained with the purely two dimensional surface theory. 
Therefore, for the rest of this chapter we will rely mostly on the effective Hamiltonian Eq.~\eqref{H_wire}.

\subsection{Aharonov-Bohm effect and magnetoconductance oscillations}
\label{sec:bands_normal_AB}

\begin{figure}[t]
\begin{center}
\includegraphics[width=6cm]{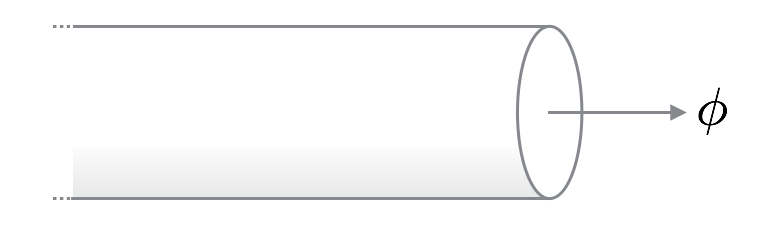}
\caption{A cylindrical topological insulator nanowire threaded by a coaxial magnetic field, resulting in total flux $\phi$ through the wire's cross-section. }
\label{flux_wire}
\end{center}
\end{figure}

\begin{figure}[t]
\begin{center}
\includegraphics[width=10cm]{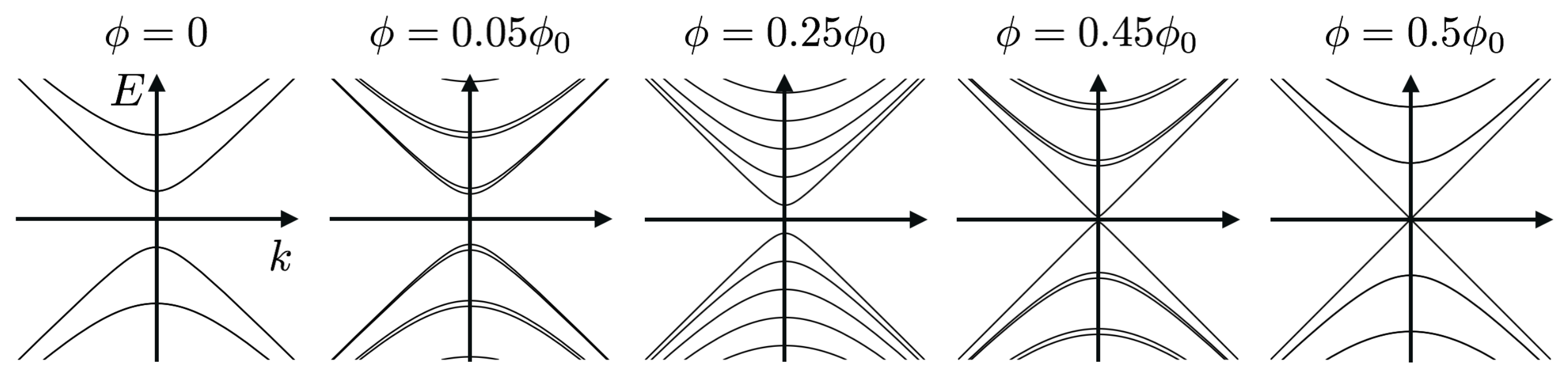}
\caption{Schematic band-structure of a topological insulator nanowire threaded by flux. 
The spectrum is composed of discrete energy bands with a given angular momentum $\ell_n$ and varying as a function of the momentum $k$ along the wire. 
At zero flux the spectrum is gapped and all bands are doubly degenerate. 
At a finite non integer or half-integer flux, there is no band degeneracy and time reversal is broken at the surface.
At half integer flux, time reversal is effectively restored, and all bands, expect the linearly dispersing one, are doubly degenerate.}
\label{eta_spectrum}
\end{center}
\end{figure}

A magnetic flux $\phi$ threading the wire's cross-section, as in Fig.~\ref{flux_wire}, results in an Aharonov-Bohm phase for the surface electrons. 
The flux is included in the Hamiltonian via minimal substitution as an azimuthal vector potential
\begin{equation}
H = v_F\left[
-i\sigma_x \partial_x +  \sigma_y( -i\partial_s + \eta )2\pi/P\right],
   \label{H_wire_flux}
\end{equation}
where $\eta=\phi/\phi_0$ is the number of flux quanta $\phi_0 = h/e$ through the cross-section. 
By a gauge transformation the flux can alternatively be absorbed into the boundary conditions as an Aharonov-Bohm phase: $\psi(s+P)= -e^{2\pi\eta}\psi(s)$.
The spectrum becomes $\eta$ dependent
\begin{equation}\label{flux_eta}
E_{k,n}(\eta)=\pm v_F\sqrt{k^2+(2\pi/P)^2(\ell_n+\eta)^2},
\end{equation}
and is shown in Fig.~\ref{eta_spectrum} for different values of $\eta$. 
By construction, the spectrum is periodic in $\eta$ and repeats whenever $\eta$ changes by an integer. 

For generic values of $\eta$ all bands are nondegenerate.
At integer and half-integer values, instead, all nonlinear bands are degenerate; in the integer case, crucially, there is a single additional nondegenerate linearly-dispersing band, corresponding to the value of $\ell_n$ for which $\ell_n+\eta=0$.
At a fixed chemical potential $\mu$, the number of modes at the fermi energy can therefore be modified by tuning the flux. 
At integer values of $\eta$ this number is always even, while at half-integer values it is always odd. 
The difference $\Delta N = N(\eta = 0) - N(\eta = 1/2)$ in number of modes is $\Delta N = \pm 1$, with the sign depending on the value of the chemical potential; at the Dirac point $\Delta N = -1$.

\begin{figure}[t]
\begin{center}
\includegraphics[width=0.75\textwidth]{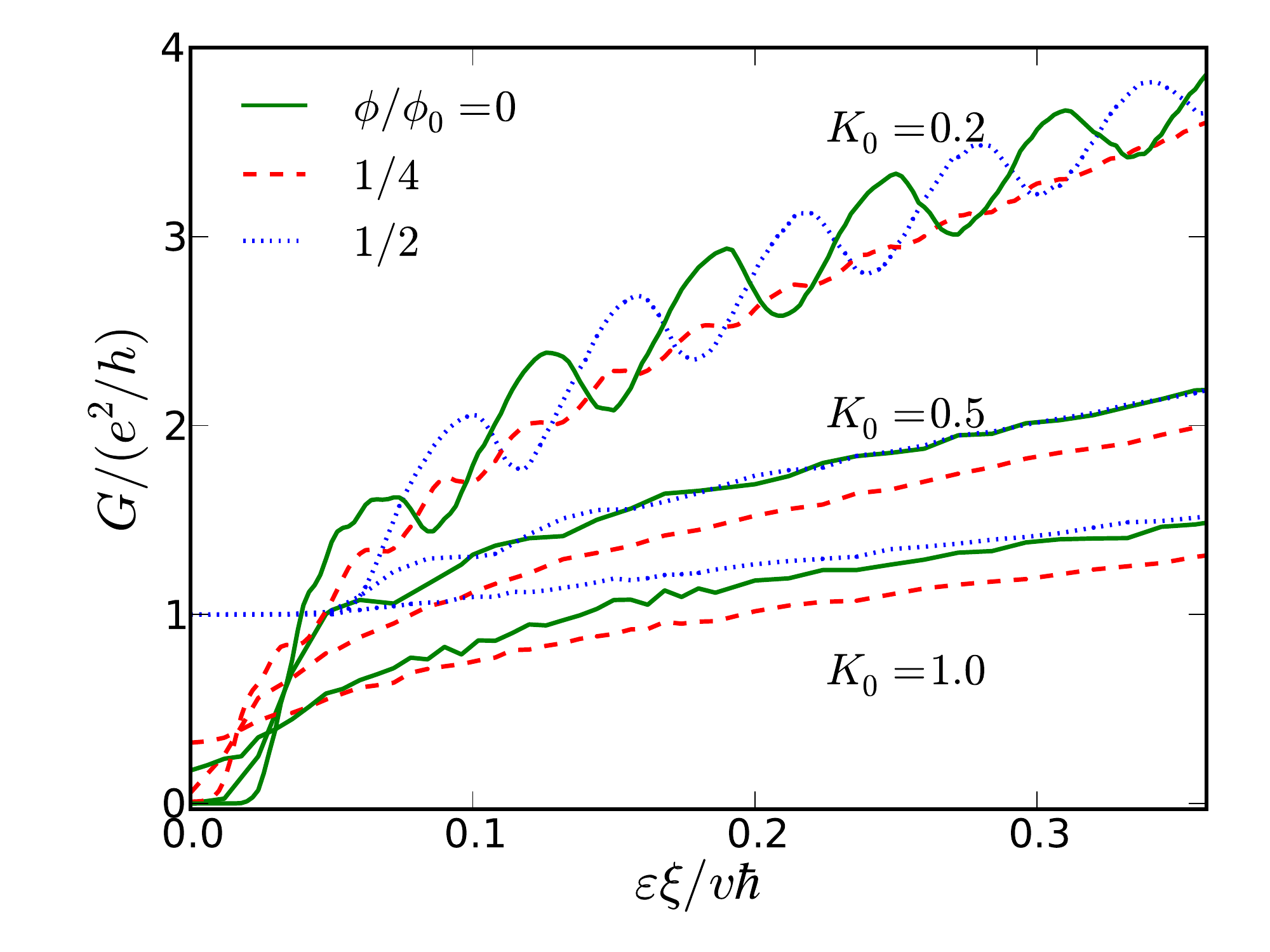}
\caption{Conductance of a topological insulator nanowire as a function of chemical potential (here denote by $\epsilon$) for three different values of flux $\phi$ and three values of the disorder strength $K_0$. The disorder is Gaussian distributed $\langle V(\mathbf{x})V(\mathbf{x}^\prime) \rangle = K_0 (\hbar v_F)^2/(2\pi\xi^2)\exp(-|\mathbf{x}-\mathbf{x}^\prime|^2/2\xi^2)$, with $\xi$ the disorder correlation length. The circumference of the wire was taken to be $P = 100\xi$ and the length $L = 200\xi$. Figure taken from Ref.~\cite{Bardarson:2010jl}}
\label{fig:AB}
\end{center}
\end{figure}

In a perfectly ballistic wire, the two terminal conductance, according to the Landauer equation, is proportional to the number of modes, $G = (e^2/h) N(\mu,\eta)$.
The above considerations then suggest that one should observe Aharonov-Bohm oscillations in the conductance with a period $\Delta \eta = 1$, corresponding to a flux periodicity of $\Delta \phi = \phi_0$, and amplitude $\Delta G = \pm e^2/h$, with a chemical potential dependent sign.
Real wires are never perfectly ballistic as there is always some amount of disorder present. 
However, as long as the disorder induced level broadening $\Gamma$ is small compared with the level spacing, $\Gamma < \Delta_P$, the above expectation should hold, with the only modification a reduced amplitude $\Delta G$ of the oscillations. 
This is borne out in numerical calculations~\cite{Bardarson:2010jl}, which modeled disorder by including a scalar potential $V(\mathbf{x})$ in the Hamiltonian~\eqref{H_wire_flux} and solving for the scattering matrix using the transfer matrix technique described in section~\ref{sec:transfer}, the results of which are shown in Fig.~\ref{fig:AB}.
At chemical potentials away from the Dirac point and at weak disorder ($K_0 = 0.2$) a clear $\phi_0$-periodic oscillations with a chemical potential dependent sign---determined by wether the blue dotted curve of $\eta = 0.5$ or green solid curve at $\eta = 0$ is higher---are clearly seen.
The transport at the Dirac point is dominated by a perfectly transmitted mode discussed in the next section.

In the opposite limit, $\Gamma \gg \Delta_P$, of strongly disordered wires, the discrete structure of the number of modes is replaced by a smoothly increasing density of states.
The conductance is no longer given by the simple mode counting argument.
To understand the flux dependence of the conductance we need to consider the symmetries of the Hamiltonian~\eqref{H_wire_flux}.
Away from integer and half-integer values of $\eta$ the time reversal symmetry $T = i\sigma_y K$, with $K$ complex conjugation, is manifestly broken by the $\eta$ term. 
However, time-reversal reemerges at half-integral and integral values of $\eta$.
This is best seen in the representation where we have gauged the flux into the boundary condition $\psi(s+P)=e^{i(2\pi\eta+\pi)}\psi(s)$; the corresponding Hamiltonian is $\eta$ independent.
The boundary conditions break time-reversal symmetry, except when $\psi(s+P)=\pm\psi(s)$, corresponding to integer or half-integer values of $\eta$.
In particular, at $\eta = 1/2$, the boundary conditions are periodic allowing for a solution with zero angular momentum and no gap---the linearly dispersing mode.

With strongly overlapping modes, $\Gamma \gg \Delta_P$, and large enough chemical potential such that $G > e^2/h$, the flux dependence of the conductance is determined by weak anti-localization.
The conductance in the presence of time-reversal symmetry is enhanced compared with that in the absence of time-reversal symmetry, due to destructive interference between time-reversed loops.
Since both half-integer and integer values of $\eta$ result in time-reversal symmetry, the period of the flux dependent conductance is $\Delta \eta = 1/2$ corresponding to flux period of $\phi_0/2$ (see $K_0=0.5$ and $K=1.0$ curves in Fig.~\ref{fig:AB})---the period is half as large as in the regime of weakly coupled modes.

There is ample experimental evidence for the weakly-coupled-mode regime being realized in current topological insulator nanowires~\cite{Peng:2010jm,Xiu:2011hq,Hamdou:2013hb,Dufouleur2013a,Hong2014,Cho:2015gk,Jauregui:2016cx}.
The magnetoconductance is found to oscillate with a period of $\phi_0$ with an amplitude whose sign can be changed by gating.
This remains true even when there is a significant bulk contribution to the conductance.
The same period is found in the flux dependence of conductance fluctuations~\cite{Dufouleur:2017de}.

We have assumed in our discussion uniformly doped wires such that the chemical potential $\mu$ is constant. 
In addition to random variations, the chemical potential can have smooth variations due to the experimental setup. 
For example, the top and bottom part of the wire may have different charge density due to the presence of a substrate.
This was studied theoretically and experimentally in the case of HgTe nanowires in Ref.~\cite{Ziegler:2018ke}.

\subsection{Perfectly transmitted mode}

The combination of time reversal symmetry and an odd number of modes, obtained at a half integral flux $\eta$, implies the existence of a perfectly transmitted mode.
The two terminal scattering matrix is antisymmetric $S^T = -S$ and the eigenvalues of the transmission matrix come in degenerate pairs~\cite{Bardarson:2008jk}; in the case of an odd number of modes one eigenvalue is exactly unity---the perfectly transmitted mode.
At the Dirac point the conductance at half integral flux is therefore quantized at $e^2/h$, irrespective of the strength or type of disorder, as long as it respects time reversal symmetry.
Away from half integral flux the conductance in the ballistic limit drops to zero as a Lorentzian with a peak width $\delta \eta = P/\pi L$~\cite{Tworzydio:2006hw}, while disorder enhances the conductance~\cite{Bardarson:2007iu,Bardarson:2010jl}.
This is evident in the numerical data of Fig.~\ref{fig:AB} where the conductance at $\eta = 0.5$ goes to $e^2/h$ at the Dirac point for all values of disorder.

These transport signatures of the perfectly transmitted mode are unique to topological insulator nanowires.
The first theoretical realization of a perfectly transmitted mode was in carbon nanotubes, where an effective symplectic time reversal symmetry is obtained in the absence of intervalley scattering and trigonal warping~\cite{Ando:2002es}.
Due to fermion doubling~\cite{Nielsen:1981ea,Nielsen:1981ke}, however, the perfectly transmitted modes always come in pairs, and the conductance therefore would be quantized at multiples of $2e^2/h$ instead of the $e^2/h$ that characterizes the topological insulators.
Furthermore, the emergent symmetry in the carbon nanotubes is easily broken and the magnetic field strengths needed to obtain a flux of $\eta = 0.5$ are huge due to the small radius of carbon nanotubes.

The effective time reversal at $\eta = 0.5$ requires a constant flux through the wire. 
Variations in the wire circumference lead to local variations in the flux, breaking the time reversal symmetry.
Random surface ripples combined with disorder result in a reduction of the conductance at the Dirac point that is no longer generally quantized~\cite{Xypakis:2017uo}.
At larger chemical potentials the amplitude of the Aharonov Bohm oscillations of the conductance reduce with increasing magnetic flux, and can, in the case of large surface ripples completely wash out the oscillations.
Experimentally realized wires can be made with a uniform enough surfaces that this effect is small.

In the absence of disorder the Hamiltonian~\eqref{H_wire_flux} at $\mu=0$ has a chiral symmetry $\sigma_z H \sigma_z = -H$, which is not broken by the random ripples in the surface.
This symmetry places the wire in the AIII symmetry class, which has a $\mathbb{Z}$ topological classification in one dimension.
The flux $\phi$ tunes the wire between topologically distinct insulating states.
At the transition between two such phases the sign of one reflection eigenvalue changes sign, requiring a perfect transmission and quantized conductance~\cite{Fulga:2011dv,Xypakis:2017uo}.

\subsection{Wires in a perpendicular field: chiral transport }\label{sec:chiral}

Apart from the perfectly transmitted mode, there exist other ways in which topologically protected chiral transport can emerge in topological insulator wires or films, which do not rely on time reversal symmetry, but rather on breaking it.
This can be achieved either by coupling the system to magnetism to induce quantum anomalous Hall phase, or by subjecting the system to a strong perpendicular magnetic field inducing the quantum Hall effect. 
In the latter case, the field is applied perpendicular to the TI surface rather than parallel as in the case of threading flux through the wire. 
Therefore, the surface state is gapped, rotational symmetry is explicitly broken, and the discussion does not benefit from considering the cylindrical geometry, but rather is easier to carry out for a rectangular wire. 

\begin{figure}[t]
\begin{center}
\includegraphics[width=10cm]{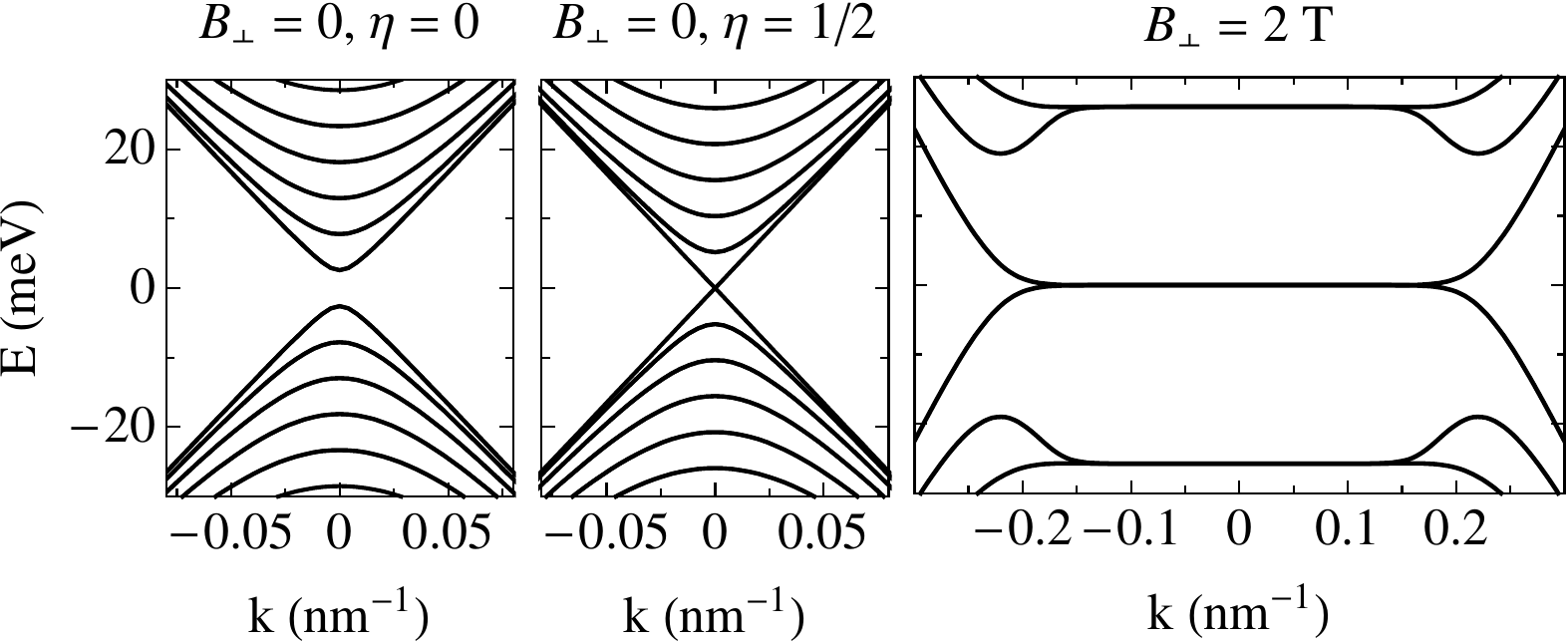}
\caption{The spectrum of a wire of cross-section $40$ nm by
 $160$ nm with and without perpendicular field $B_\perp$, with and without a vortex along the wire, as a function of the momentum along the wire. Note that in the last case when $B_\perp$ is nonzero the spectrum does not depend on the presence of absence of the vortex. Figure taken from Ref.~\cite{DeJuan:2014fs}}
\label{B_spectrum}
\end{center}
\end{figure}

A strong field applied perpendicular to a TI rectangular wire breaks the spectrum of the surface states into Landau levels with unique characteristics stemming both from the Dirac like behavior of the particles, as well as from the fact that the surface has no boundary~\cite{Lee:2009do}. 
While the top and bottom surfaces are gapped, the side surfaces, which are parallel to the direction of the external field, remain gapless. 
In the absence of termination of the surface state, chiral quantum Hall edge states, analogous to those resulting from the presence of a confining potential in purely two dimensional systems, exist on the sides surfaces, see Fig.~\ref{B_spectrum}. 

A single Dirac fermion in a flat infinite space realizes the half-integer quantum Hall effect with $\nu = n+1/2$~\cite{Brey:2014db}.
In the case of a compact surface with uniform doping, the top and bottom surface have the magnetic field pointing in the opposite direction compared with the surface normal, and are therefore in the opposite quantum Hall state: if the upper surface is in state $\nu$, the lower is in state $\nu^\prime=-\nu$~\cite{Lee:2009do}.
This results in quantum Hall plateaus with $\Delta \nu = \nu - \nu^\prime = 2n+1$ and an associated two terminal conductance of $\sigma_{xx} = (2n+1)e^2/h$.
The side surfaces therefore host an odd number of chiral edge states. 
The Hall conductivity $\sigma_{xy}$ depends on the detailed configuration of current and voltage leads~\cite{Sitte:2012ib,Konig:2014gb}.
In principle, the doping of the top and bottom surface can be tuned separately, realising all integer states $\nu = n + n^\prime + 1$, including $\nu = 0$~\cite{Brune:2011hi,Feng:2015jq,Morimoto:2015cy}.

The Landau level at the charge neutrality point is special in that it is a combination of electron-like and a hole-like Landau levels, while all higher or lower Landau levels are purely electron-like or hole-like.
At charge neutrality, counter-propagating modes are therefore found close in both energy and space, allowing scalar disorder to couple them such that a $\nu=0$ quantum Hall plateau is obtained in a narrow window of energies~\cite{Xypakis:2017kw}. 
The width of this window increases with increasing disorder strength.
This $\nu = 0$ state is distinct from the one induced by doping the top and bottom surfaces differently.

The chiral edges states can be probed in various ways; here we mention three.
First, In the lowest Landau level regime, only a single chiral edge state moves on each side surface, and they move in opposite direction on each surface.
The direction in which the chiral moves depends on the doping. 
In a p-n junction, therefore, on each surface one obtains counter-propagating modes that meet in the transition region between the p and n halves of the junction.
Since they can not disappear they instead travel along the p-n junction interface to the other side of the wire, where they can propagate away and into the lead.
The obtained conductance depends on the overlap of the spin of chiral states and the phase they pick up while crossing the junction; this latter phase can be controlled by a flux along the wire, realizing a Mach-Zender interferometer~\cite{Ilan:2015ei}.

Second, the higher Landau levels have a characteristic non-monotonic dependence on the longitudinal momentum: the degenerate Landau levels are spilt as they turn into edge states and one of them dips in energy below the energy of the Landau level.
This non-monotonic dispersion has a surprising effect on thermal transport.
Namely, when applying the right temperature difference between two leads, one can obtain a particle current flowing from the cold reservoir to the hot, counter to intuition~\cite{Erlingsson:2017bj}.

Finally, the single chiral mode limit is useful in probing topological superconductivity, since in this case the two terminal conductance becomes a direct probe of the presence of Majorana modes~\cite{DeJuan:2014fs}. 
Superconducting proximity effect and transport is the subject of the next section.

\section{Topological Insulator nanowires and superconductivity}
\label{sec:3}

Proximity induced superconductivity in materials with spin-orbit coupling is one of the promising schemes to engineer superconducting states with non-trivial topological properties~\cite{Beenakker:2013jb,Alicea:2012hz}. 
One of the first proposals for such an engineered phase is a topological insulator put in proximity with an s-wave superconductor~\cite{Fu:2008gu,Fu:2009hd}. Such a construction is predicted to yield a one-dimensional topological superconductor at the edge of a quantum spin Hall sample, or a two dimensional topological superconductor at the surface of a three dimensional topological insulator. 

Other prominent examples are one dimensional nanowires made from materials such and InAs under the application of an external magnetic field, or magnetic chains, which under proximity effect can form an effective p-wave superconductor in one dimension~\cite{Lutchyn:2010hp,Oreg:2010gk}. Recently, such systems have shown signatures consistent with the appearance of zero energy modes at their ends, a central characterizing feature of topological superconductors in one dimension~\cite{Mourik:2012je,Das:2012hi,NadjPerge:2014ey}. This was shown both through transport as well as in Scanning Tunneling Microscopy. 

In this section we review how TI in three dimensions formed into nanowires represent a novel and tunable version of a quasi-one dimensional topological superconductor, using the elements described in previous section. To this end, we begin by recalling the essential requirements for a normal one dimensional system to become a topological superconductor under proximity effect.

\subsection{Topological superconducting phases in one dimension}
Almost two decades ago, Kitaev put forward simple criteria for the emergence of topological superconductivity in one dimension, and formulated a topological invariant, which can be calculated from the lattice model representing it, which determines the fate of the superconducting phase~\cite{Kitaev:2001gb}. Essentially, the main criteria require that the underlying normal system has an odd number of Fermi points in the right half of the Brillouin zone, as well as a finite gap when superconductivity is introduced. In light of the discussion above, TI nanowires become immediate suspects for becoming topological superconductors in one dimension when pierced with one half of a flux quantum.   

The topological invariant characterizing the one dimensional lattice system, known as the Majorana number, is most generally defined as 
\begin{equation}
\mathcal{M}=\text{sgn}\left[\text{Pf}\tilde{B}(k=0)\right]\text{sgn}\left[\text{Pf}\tilde{B}(k=\pi)\right]
\end{equation}
Here, $\text{Pf}$ stands for Pfaffian, and  $\tilde{B}$ is the Hamiltonian matrix  of a (quasi-) one dimensional system expressed in a Majorana basis. In the limit of small pairing potential $\Delta$, this expression reduces to a much simpler one:
\begin{equation}
\mathcal{M}=(-1)^\nu
\end{equation}
where $\nu$ is an integer counting the number of Fermi points.  A non-trivial phase is labelled by $\mathcal{M}=-1$, and is expected to have zero modes when surfaces are introduced. 

The first to consider this topological invariant in the context of TI nanowires were Cook and Franz~\cite{Cook:2011ij}, predicting that a cylindrical nanowire combined with superconductivity is expected to have a non-trivial Majorana number when the flux through the wire is close to $\pi$. 

The Hamiltonian of a cylindrical wire in the presence of a pairing potential induced by proximity to an s-wave superconductor is given by
\begin{equation}
H^{(n_v)} = \left[
-i\sigma_x \partial_x +  \sigma_y( -i\partial_s + \eta \;\tau_z)2\pi/P - \mu 
\right]\tau_z + \Delta_0   \theta(-x) e^{-i \tau_z n_v s}\tau_x , \label{BdGv}
\end{equation}
where $\Delta_0$ is the superconducting pair potential induced by an adjacent bulk s-wave superconductor, assumed to be a constant. The phase of the order parameter represents an important degree of freedom and is allowed to wind around the wire. This winding has been explicitly singled out here in the exponential factor $e^{-i \tau_z n_v s}$, where $n_v$ denotes the number of vortices with a  core that is co-aligned with the wire's axis. 

\begin{figure}[t]
\begin{center}
\includegraphics[width=4.5cm]{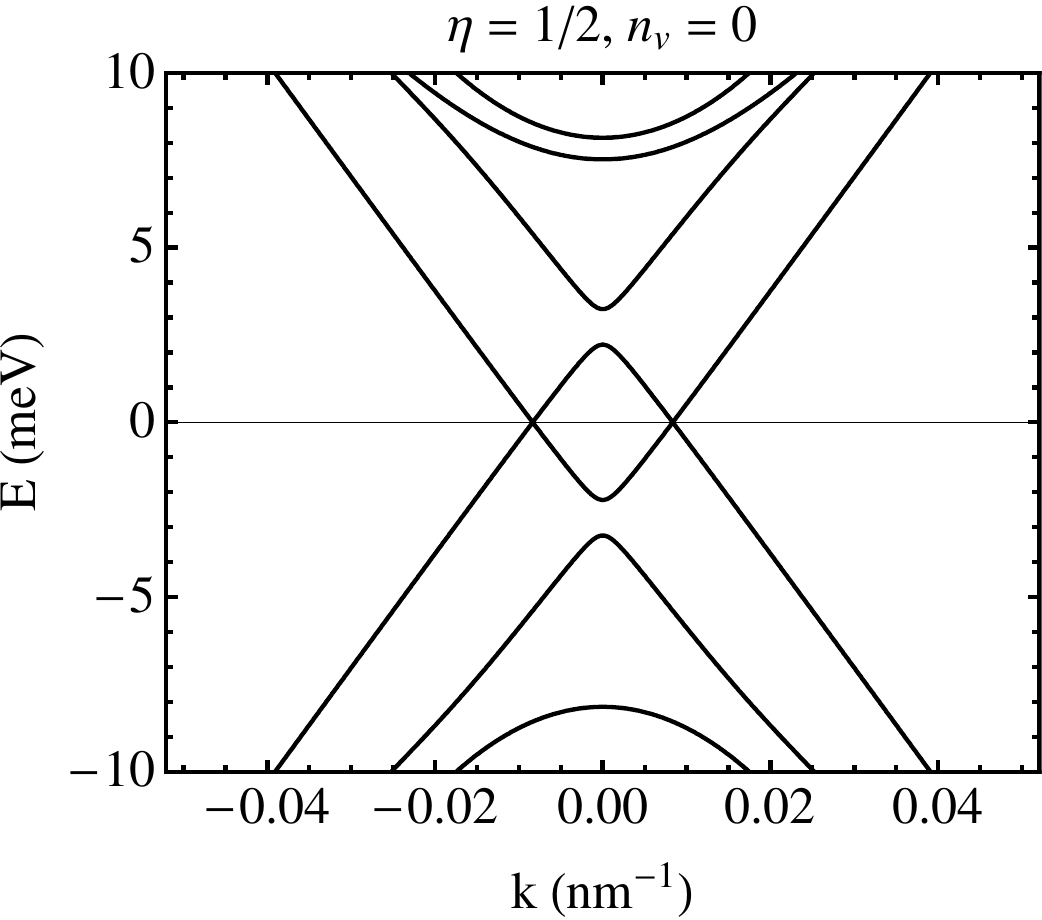}\includegraphics[width=4.5cm]{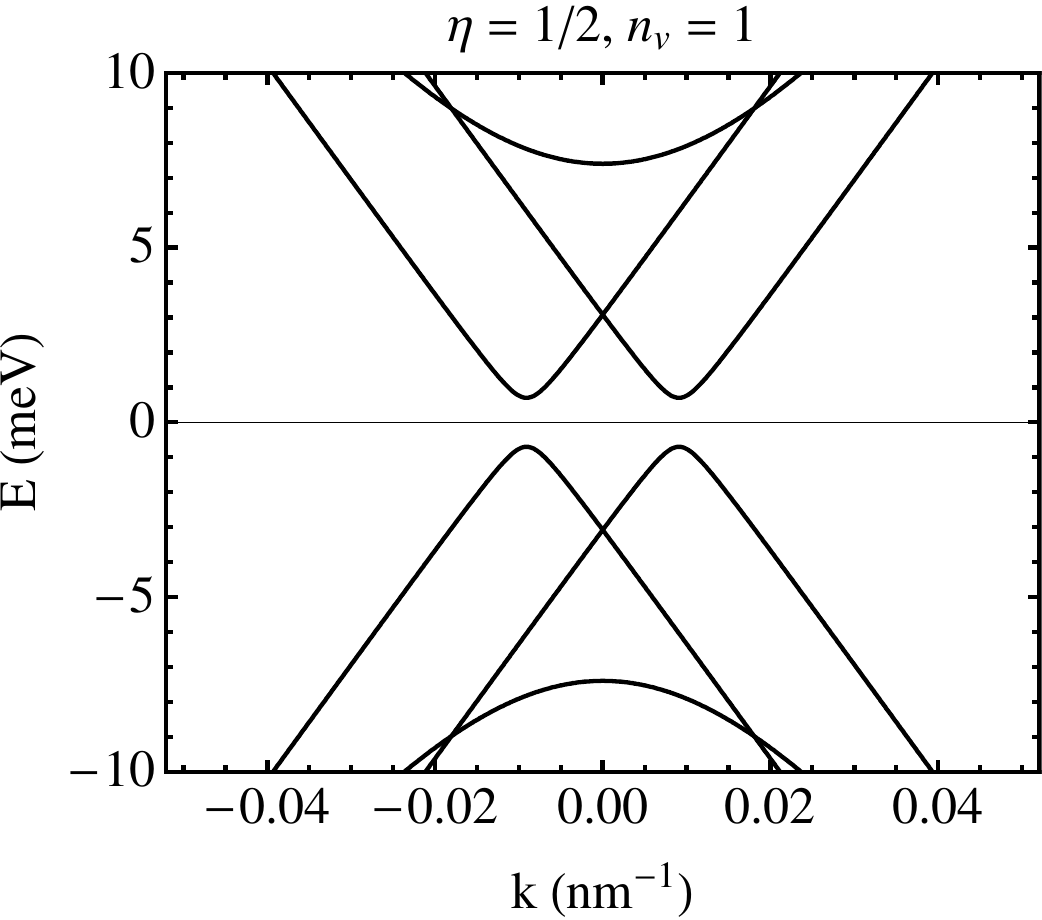}
\caption{Spectrum of the BdG Hamiltonian Eq.~\eqref{BdGv} describing the surface state of a cylindrical TI nanowires with proximity induced superconductivity. The left panel is for $n_v=0$, and the spectrum is gapless. The right panel represent the same spectrum, for $n_v=1$, and is clearly gapped.}\label{spec_sc}
\end{center}
\end{figure}

The importance of phase winding in cylindrical wires was stressed in Ref.~\cite{DeJuan:2014fs}, and can be easily argued by considering the band-structure of the wire as the spectrum of Eq.~\eqref{BdGv} with and without a vortex, i.e, the difference between $n_v=0$ and $n_v=1$.  The full characterization of the phase diagram of the wire is obtained by considering, in parallel, the topological invariant, and the energy gap in the spectrum. In order to obtain a topological superconductor, two conditions must be met simultaneously: the spectrum must have a finite gap, and the Majorana number must be equal to $-1$. As is evident from Fig.~\ref{spec_sc}, in the absence of a vortex, the spectrum is gapless, although a calculation of the Majorana number will yield a non-trivial value. 

To understand the role of the vortex, we remind ourselves that an s-wave pairing potential couples fermion states to form cooper pairs of zero total angular momentum and spin at the Fermi energy. In order for this pairing to become effective, energy bands of particles and holes in the BdG spectrum must cross at the Fermi energy with the appropriate quantum numbers. Considering the normal state band-structure discussed in sections~\ref{sec:bands_normal} and~\ref{sec:bands_normal_AB}, we note that the energy bands crossing at the fermi energy have a mismatch of angular momentum, hence s-wave pairing cannot open a gap at the chemical potential. However, if a winding of the order parameter is introduced, it can act to compensate for the mismatch of angular momentum, enabling the opening of a finite gap. Note, however, that this sharp statement strongly relies on the rotational symmetry of the problem, and may soften when this symmetry is structurally violated.

\subsection{Boundaries and interferences: zero modes}

\begin{figure}[t]
\begin{center}
\includegraphics[width=10cm]{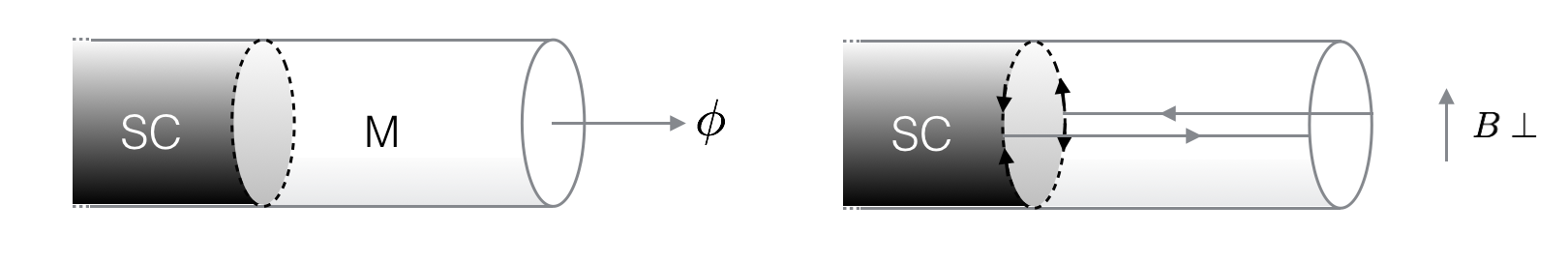}
\caption{An interface at the surface of a TI nanowire. On the left: boundary between superconductivity and magnetism traps a Majorana zero mode in the presence of flux. On the right: boundary between a gapped quantum Hall phase and a SC phase. }\label{SC_M_inter}
\end{center}
\end{figure}

\begin{figure}[t]
\begin{center}
\includegraphics[width=6cm]{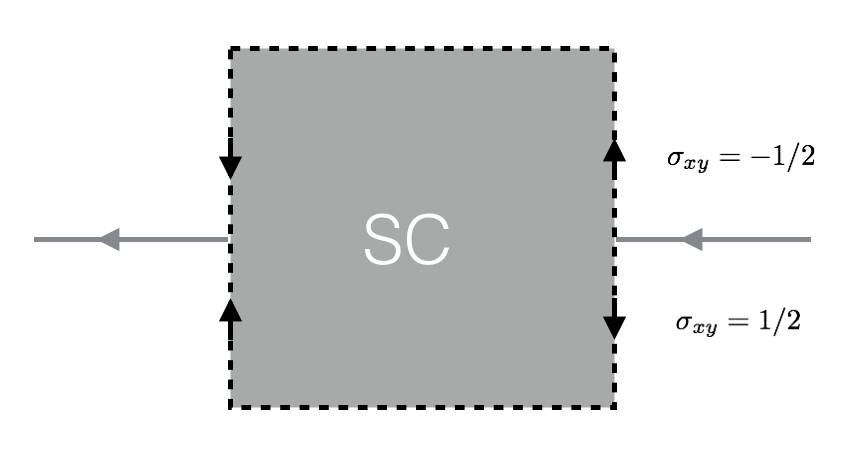}
\caption{Unfolded geometry of the TI nanowire with a normal-superconducting interface in a perpendicular field. }
\label{fig:interferometer}
\end{center}
\end{figure}

Once the emergence of topological superconductivity is well established, two questions immediately arise. The first concerns the fate of boundary states, and the second concerns signatures of them on experimentally observable quantities. It is well known that topological superconductors support Majorana edge or boundary modes, and their presence at the surface of topological insulators is no exception. 

The two dimensional topological superconductor formed at the surface of a proximitized three dimensional topological insulator is a well studied phase~\cite{Fu:2008gu,Qi:2011hb,Hasan:2010ku}. It crucially differs from the prototypical two dimensional topological superconductor in two dimensions, the p-wave superconductor, by the fact that it respects time reversal symmetry. It also differs from it by the fact that it has no natural boundary---the surface cannot terminate. In order to gain access to zero modes and boundary states, it is therefore necessary to interface a region with a topologically non-trivial superconducting gap with another region that has a different topological gap in order to trap gapless boundary states, or else create a topological defect such as a vortex or a Josephson junction. 

The two mechanisms that can gap out the surface states of topological insulators are breaking charge conservation and time reversal symmetry. Hence, it is expected that when two regions gapped by these are interfaced, gapless modes should arise~\cite{Fu:2008gu,Fu:2009hm,Akhmerov:2009jp}. Breaking time reversal symmetry can be obtained in two ways: either by subjecting the system to a magnetic field that couples to the orbital motion of the particles, as discussed extensively in Sec.~\ref{sec:chiral}, or by magnetically doping the material and introducing an energy gap via Zeeman coupling to the particle's spin. Both mechanisms have been considered in the context of wires as ways to trap Majorana bound states. 

Ref.~\cite{Cook:2011ij} considered a magnetic domain interfaced with superconductivity at the surface of a TI nanowire, following the realization of Fu and Kane that such an interface will host a chiral Majorana mode whose direction of propagation depends on the sign of the magnetization. In the case of a nanowire, this mode will form a compact loop around the wire, and therefore will have a discrete energy spectrum, tunable via the boundary conditions. When half of a flux quantum threads the wire, this Majorana mode might have zero angular momentum and therefore a zero mode in its spectrum, which is protected since its counterpart is spatially separated from it and resides at the other end of the superconducting domain. Its wave function is exponentially localized in both regions over a length scale that is set by the two gaps: $\ell_\Delta=\Delta_0/v_F, \ell_m=|m|/v_F$, where $|m|$ is the amplitude of the magnetization. 

As reviewed in Sec.~\ref{sec:chiral}, a magnetic field perpendicular to the surface of the TI, introduces Dirac-like Landau levels, with the lowest one contributing a Hall conductance of $\sigma_{xy}=e^2/2h$, which one can understand as stemming from the chiral modes living on the side surfaces. Each fermionic chiral mode  is predicted to be broken into two Majorana fermion chiral modes in the presence of superconductivity, since quite generically, chiral fermionic mode can always be trivially written as a superposition of two chiral co-propagating one dimensional Majorana modes. In the absence of superconductivity, these are constrained to move together, a constraint that can only be removed in the presence of a pair potential. 

The nanowire geometry in principle should allow to spatially separate these modes~\cite{DeJuan:2014fs}. Consider a wire with normal-superconducting interface at some point $x_0$ along the wire. When the wire is subjected to a strong magnetic field (which we assume is fully screened in the superconducting region), chiral modes will flow on the side surfaces of the normal region to and from the normal-superconducting interface, as depicted in Fig.~\ref{SC_M_inter}. At the interface, the chiral fermion modes are broken into two Majorana modes, that flow in the top and bottom surfaces of the wire, along the interface. These modes break apart on one side surface and recombine on the other, forming a Majorana interferometer, see Fig.~\ref{fig:interferometer}. The relative phase for the two chiral Majorana modes that encircle the wire can be tuned by introducing a vortex along the wire's axis.

\subsection{Transport signatures of topological superconductivity}
\begin{figure}[t]
\begin{center}
\includegraphics[width=5cm]{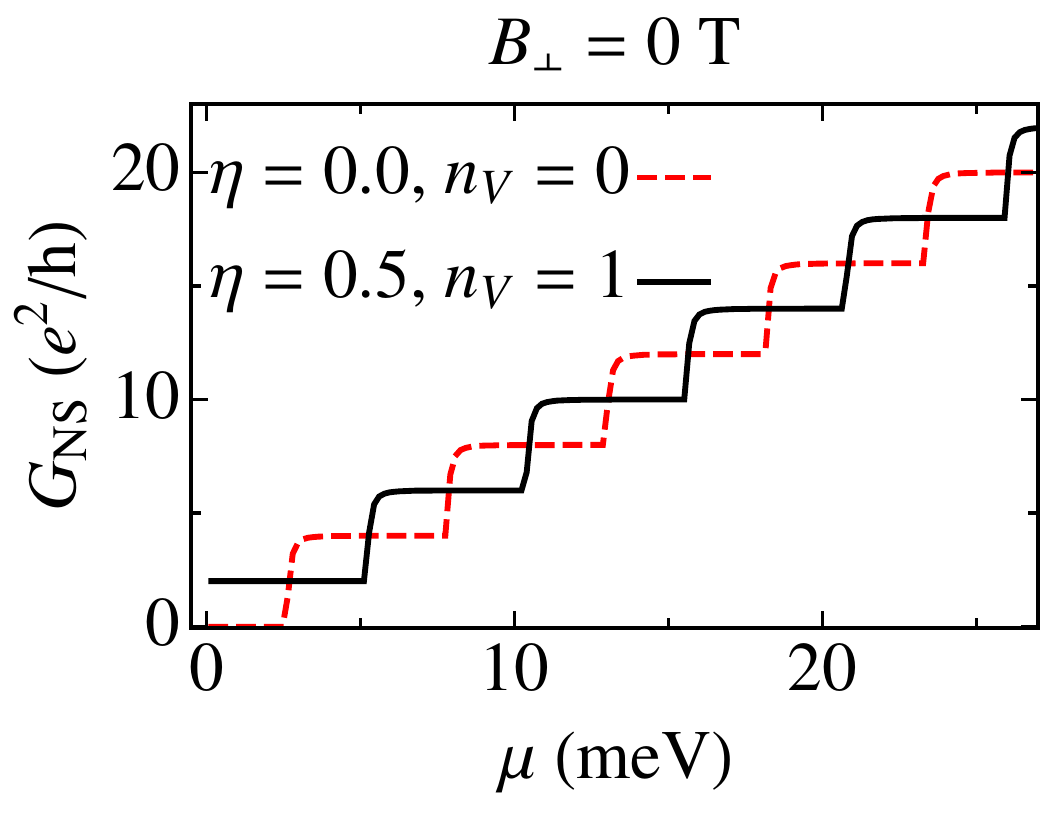}\includegraphics[width=5cm]{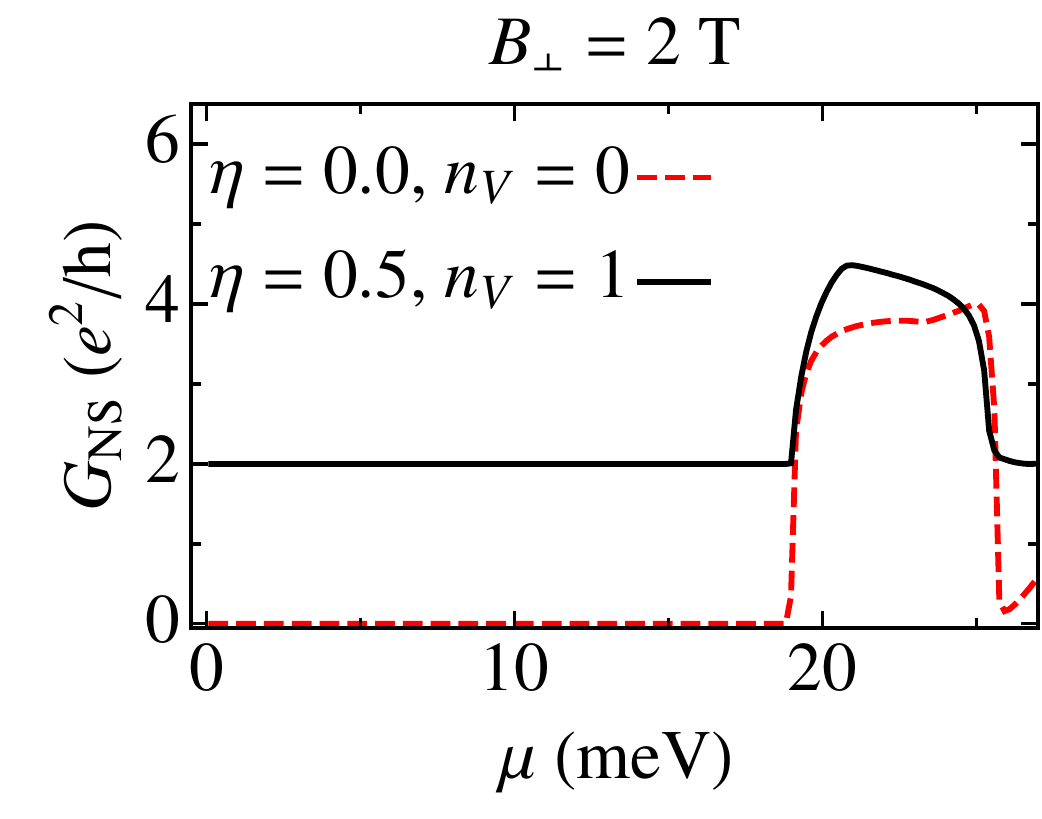}
\caption{The two terminal conductance across a normal superconducting interface with (right) and without (left) a perpendicular magnetic field, without and without a vortex along the wire.  A conductance plateau at $2e^2/h$ can clearly be seen when a vortex is present ($n_v=1$) in both cases, but the plateaux is wider with the perpendicular field present. Figure taken from Ref.~\cite{DeJuan:2014fs}}\label{NSconductance}
\end{center}
\end{figure}

The layout of the TI nanowire as described at the end of the previous section corresponds to a Majorana two path interferometer~\cite{DeJuan:2014fs}, where the two arms pick up a different phase that is directly correlated with the phase winding of the superconducting order parameter in the azimuthal direction (namely, around the wire), and is in one-to-one correspondence with the Majorana interferometer designed by Fu and Kane as well as Akhmerov, Nillson and Beenakker~\cite{Fu:2009hm,Akhmerov:2009jp}. However, there is a crucial difference between the these previous proposals and the one based on the nanowire geometry: realizing earlier proposals necessitates the use of magnetic domains of opposite magnetization on a single surface, as well as a magnetic field to generate phase winding for the two arms by threading vortices within a superconducting region enclosed by the interferometer's arms. The coexistence of these ingredients presents a great experimental challenge. Interfacing magnetism with a topological insulator surface state as means to create chiral modes has so far proven difficult, while small superconducting islands with a tunable number of vortices are yet to be attempted.  

The nanowire geometry seems to be a natural setting for the realization of the interferometer, one that bypasses some of the difficulties mentioned above. The small circumference of the nanowire allows the compactification of the chiral Majorana modes, while the flux through the wire enables the twisting of the boundary conditions for that mode through the formation of the vortex along the wire. It is therefore natural to look for the signatures such a mode in plain two terminal conductance along the wire, measured across the interface between the SC and the normal state wire~\cite{Wimmer:2011fq}. Indeed, theoretical predictions have been made for the two terminal conductance which predict either perfect Andreev reflection or perfect normal reflection~\cite{DeJuan:2014fs}. The conductance of the wire, presented in Fig.~\ref{NSconductance}, shows a clear hallmark of the interface Majorana mode: a flat $2e^2/h$ conductance plateau appears in the presence of a vortex along the wires core, which can be enhanced by the application of an external magnetic field. We stress that the perpendicular field is not a prerequisite for such a plateau to appear, but its presence enhances the chemical potential range in which the system is in the single mode regime, as demonstrated in Fig.~\ref{B_spectrum}, and required for topological superconductivity to appear. 

Additional signatures of topological superconductivity in nanowires are predicted to appear in transport across Josephson junctions~\cite{Alicea:2012hz,Beenakker:2013jb,Ilan:2014ef}. There, at a phase difference of $\pi$ across the junction, a non-chiral Majorana mode is expected to be trapped within the junction provided, again, that a phase winding of $\pi$ to the order parameter is properly introduced by a co-axial flux. The appearance of Majorana modes within Josephson junctions is expected to result in a $4\pi$ periodic current phase relation, provided no stray quasi-particles induce parity switches of the low lying state in the junction. Such a universal prediction is not unique to TI wires. The current phase relation should also have a distinct skewed shape in the absence of parity conservation, but perhaps the most striking signature is expected to appear in $I_cR_N$, namely, the product of the critical current and the normal state resistance: $I_cR_N$ is expected to peak sharply when half of a flux quantum induced a vortex through the wire's core~\cite{Ilan:2014ef}.

\section{Technical details: transfer matrix technique}
\label{sec:transfer}
Assume we have a Dirac Hamiltonian of the form (in this section we set $v_F = 1$)
\begin{equation}
	H = \Gamma_x p_x + \Gamma_y p_y + \hat{V}(x,y),
	\label{eq:DiracGeneral}
\end{equation}
and want to calculate the two terminal scattering matrix $S$. 
The $2N\times 2N$ matrices $\Gamma_x$ and $\Gamma_y$  depend on the problem at hand; for the single Dirac fermion in Eq.~\eqref{H_wire} we have $\Gamma_x=\sigma_x$ and $\Gamma_y = \sigma_y$, but for weak topological insulators they are certain $4\times 4$ matrices. 
The potential term $\hat{V}$ is a matrix like the $\Gamma$'s, constrained only by symmetries.

The scattering matrix relates incoming scattering states to outgoing scattering states~\cite{Beenakker:1997gz}
\begin{equation}
	\begin{pmatrix}
		\psi^\text{out}_L \\
		\psi^\text{out}_R
	\end{pmatrix}
	=
	\begin{pmatrix}
		r & t' \\
		t & r'
	\end{pmatrix}
	\begin{pmatrix}
		\psi^\text{in}_L \\
		\psi^\text{in}_R
	\end{pmatrix}.
\label{eq:S}
\end{equation}
The transfer matrix $M$ relates the wave function at two points~\cite{Beenakker:1997gz}
\begin{equation}
	\psi(x) = M(x,x')\psi(x').
\end{equation}
In the basis of scattering states, $M = M(L,0)$ takes the form
\begin{equation}
	\begin{pmatrix}
		\psi^\text{in}_R \\
		\psi^\text{out}_R
	\end{pmatrix}
	= M
	\begin{pmatrix}
		\psi^\text{in}_L \\
		\psi^\text{out}_L
	\end{pmatrix},	
\end{equation}
and from the definition of the scattering matrix~\eqref{eq:S}, $M$ takes the form
\begin{equation}
	M = \begin{pmatrix}
	{t^\dagger}^{-1} & r'{t'}^{-1} \\
	-{t'}^{-1}r & {t'}^{-1}
	\end{pmatrix}.
\end{equation}
Therefore, from the transfer matrix we obtain the scattering matrix; the two terminal conductance is given by the Landauer formula
\begin{equation}
	G = \text{tr} \; t^\dagger t.
\end{equation}

The general Dirac equation~\eqref{eq:DiracGeneral} is transformed into the basis of scattering states by the unitary transformation $U$ satisfying $U^\dagger \Gamma_x U = \Sigma_z$, where $\Sigma_z = \sigma_z\otimes \mathbb{1}_N$ is a diagonal matrix with the first $N$ entries on the diagonal equal to 1 and the last $N$ entries equal to $-1$.
The Hamiltonian takes the form
\begin{equation}
	H = \Sigma_z p_x + \tilde{\Gamma_y} p_y + \hat{\tilde{V}}(x,y).
\end{equation}
where $\tilde{\Gamma_y} = U^\dagger \Gamma_y U$ and $\hat{\tilde{V}} = U^\dagger \hat{V} U$.

Using this form of the Hamiltonian to integrate the Dirac equation $H\psi = E\psi$, the transfer matrix takes the form 
\begin{equation}
	M = \mathcal{T}_x \exp \left\{i\Sigma_z\int_0^L \left[E - \tilde{\Gamma}_yp_y - \hat{\tilde{V}}(x,y)\right] dx\right\},
\end{equation}
where $\mathcal{T}_x$ is the position ordering operator needed if the terms in the integral do not commute at different $x$, which happens for example if $\hat{V}$ depends on $x$. 
Using the transitive property $M(x_1,x_3) = M(x_1,x_2)M(x_2,x_3)$ and separating the integral into $N_x$ equally spaced intervals, one can approximate the above expression by dropping the position ordering, which is valid if $L/N_x \ll \xi$ the correlation length of the potential $\hat{V}$, obtaining
\begin{equation}
	M = \prod_{i=1}^{N_x} \exp \left\{i\Sigma_z\int_{x_i}^{x_{i+1}} \left[E - \tilde{\Gamma}_yp_y - \hat{\tilde{V}}(x,y)\right] dx\right\} = \prod_{i=1}^{N_x} M_i,
	\label{eq:M}
\end{equation}
with $x_1 = 0$ and $x_{N_x+1} = L$.
This equation can be solved numerically in the basis of $p_y$ momentum eigenstates in which case, for a fixed value of $x$, $\hat{\tilde{V}}$ is the $N_y \times N_y$ matrix
\begin{equation}
	\hat{\tilde{V}}_{nn^{\prime}}(x) = \int_{0}^{P} \frac{dy}{ P} e^{i(q_n - q_{n^{\prime}})y} 
\hat{\tilde{V}}(\mathbf{r}).
	\label{}
\end{equation}
$N_y$ is the total number of momentum modes included in the calculation and $q_n$ are the eigenvalues of $p_y$.
In principle, since we are in the continuum, $N_y$ is infinite; in practice, however, the transmission of high transverse momentum modes is negligible and the matrix can be truncated at some cutoff momentum that is taken large enough that the conductance is independent of it. 
Similarly, $N_x$ is increased until convergence is obtained.

The transfer matrix $M$ has exponentially large and small eigenvalues and the matrix product~\eqref{eq:M} is numerically unstable.
The unitary scattering matrix in contrast, has complex eigenvalues with unit amplitude. 
It is therefore useful to use the relation between transfer and scattering matrices to transform the transfer matrix $M_i$ of the $i$-th interval into a scattering matrix $S_i$. 
The product of transfer matrices becomes a convolution of scattering matrices
\begin{equation}
	S = \bigotimes_i S_i,
\end{equation}
with the convolution defined by
\begin{equation}
         \begin{pmatrix} 
        r_1 & t_1^\prime \\
        t_1 & r_1^\prime
        \end{pmatrix} \otimes
        \begin{pmatrix} 
        r_2 & t_2^\prime \\
        t_2 & r_2^\prime
        \end{pmatrix} =
           \begin{pmatrix} 
		 r_1+t_1^\prime r_2(1-r_1^\prime r_2)^{-1} t_1 & t_1^\prime(1-r_2 r_1^\prime)^{-1} t_2^\prime \\
		 t_2(1-r_1^\prime r_2)^{-1} t_1 & r_2^\prime+ t_2 r_1^\prime
		 (1-r_2 r_1^\prime)^{-1} t_2^\prime
        \end{pmatrix}.
	\label{}
\end{equation}

As a demonstration consider the 2D Dirac Hamiltonian~\eqref{H_wire}. 
Since it is translationally invariant the position ordering can be dropped and the integral over $x$ performed trivially since the integrand is independent of $x$, resulting in the transfer matrix
\begin{equation}
	M = \exp \left[i\sigma_z \left(\mu + \sigma_y p_y \right) L \right],
\end{equation}
where we have taken $E=\mu$.
From this expression we obtain the scattering matrix. 
At the Dirac point $\mu = 0$, in particular, 
\begin{equation}
	M = U \begin{pmatrix} e^{p_yL} & 0 \\ 0 & e^{-p_yL} \end{pmatrix} U^\dagger,
\end{equation}
with $U^\dagger \sigma_x U = \sigma_z$.
This finally gives 
\begin{equation}
	t = 1/\cosh(p_y L),
\end{equation}
consistent with Ref.~\cite{Tworzydio:2006hw}.

\section{Experimental status and outlook}

Some of the theoretical aspects of the theory of the normal state transport in topological insulator nanowires presented here have already been tested, and some confirmed experimentally. Multiple groups have achieved Aharonov-Bohm interference in several nanowires of lengths ranging from hundred of nanometers to several microns, and circumference of approximately $200$ nanometers long~\cite{Peng:2010jm,Xiu:2011hq,Hamdou:2013hb,Dufouleur2013a,Hong2014,Cho:2015gk,Jauregui:2016cx,Dufouleur:2017de}. Oscillations were observed in magnetic fields equivalent to up to $10$ flux quanta threaded through the cross section of the wire, and signatures consistent with the emergence of a perfectly transmitted mode were also observed, showing up as an enhances conductivity at half integer flux quanta threaded through the wire. 

In addition, 3DTI nanowires were also coupled to superconductivity. The Josephson effect was recently measured in $\text{BiSbTeSe}_2$ wires coupled to superconducting Niobium leads, displaying an anomalous behavior indicating the formation of low energy Andreev bound states at the crossover from short junction to long junction behavior~\cite{Kayyalha:2017tk}. The ability to resolve low energy modes is a promising step towards the realization and detection of Majorana bound states in such junctions. Nevertheless, it has been suggested that additional physics related to the Kondo effect might emerge in the presence of normal-superconducting interfaces~\cite{Cho:2016ft}, alluding to a different origin for the emergence of zero bias peaks in transport across such an interface. This certainly calls for additional exploration of transport in 3DTI nanowire based heterostructure, both theoretically and experimentally. 

Finally, the prospect of using 3DTI wires as a competitive platform for topological quantum computation is still being explored. A recent work has proposed an architecture made from  coupled 3DTI nanowires based Majorana box qubits, namely short segments of proximitized 3DTI wires connected by gapped 3DTI normal wire segments, as means to implement simple quantum operations on single qubits~\cite{Manousakis:2017kk}. It will be both interesting as well as a challenge to bring such architectures to life both from the materials perspectives, as well as conceptually bridging the gap between these novel ideas and the limitations of the actual experimental system.

\begin{acknowledgement}
We thank Fernando de Juan and Joel Moore for collaborations, and Yong P. Chen for multiple discussion regarding the experimental systems. We would also like to thank Fernando de Juan for contributing figure ~\ref{spec_sc} to this review. Work on this review was supported by the ERC Starting Grant No. 679722 and the Knut and Alice Wallenberg Foundation 2013-0093.
\end{acknowledgement}

\bibliographystyle{unsrt}
\bibliography{references}

\end{document}